\documentclass[11pt]{article}
\usepackage{styleFile}
\title{A Stochastic Operator Approach to Model Inadequacy with Applications to Contaminant Transport}
\author{T. Portone, D. McDougall, R. Moser}

\begin{document}
\maketitle
\begin{abstract}
\pagenumbering{gobble}
The mathematical models used to represent physical phenomena are generally known to be imperfect representations of reality.
Model inadequacies arise for numerous reasons, such as incomplete knowledge of the phenomena or computational intractability of more accurate models.
In such situations it is impractical or impossible to improve the model, but necessity requires its use to make predictions.
With this in mind, it is important to represent the uncertainty that a model's inadequacy causes in its predictions, as neglecting to do so can cause overconfidence in its accuracy.
A powerful approach to addressing model inadequacy leverages the composite nature of physical models by enriching a flawed embedded closure model with a stochastic error representation.
This work outlines steps in the development of a stochastic operator as an inadequacy representation by establishing the framework for inferring an infinite-dimensional operator and by introducing a novel method for interrogating available high-fidelity models to learn about modeling error.

\end{abstract}
\pagenumbering{arabic}

\section{Introduction}
\subsection{Predictive validation framework}
The use of models that are known to be imperfect is common in simulating complex systems. 
There are many possible causes for such imperfections:
model inputs such as initial and boundary conditions, geometry, and model parameters are often unknown;
there is an incomplete knowledge of the physics involved; 
or, if there is, accurate models are computationally intractable. 
Additionally, observational data is often sparse and uncertain.
One must therefore ask how these inadequate models can be used with confidence to make predictions that will guide crucial decisions.
These predictions generally require the model to extrapolate to scenarios in which observations of quantities of interest (QoIs) are unavailable, making it difficult to assess the accuracy of the predictions. 
This makes understanding the effect of all uncertainties crucial in the assessment of the model's reliability.

Validating a model for prediction requires a more nuanced approach than simply comparing its solution to data.
There is no direct way to test the model's predictions, so confidence in its ability to extrapolate must be built up by testing the model and its component parts in increasingly challenging ways.
An important feature of physics-based models that enables their use in predictions is that they are a composition of highly reliable theories, such as conservation laws, and less reliable ``embedded models'' that are needed to close the model.
Oliver et al. leverage this structure to better characterize a model's reliability and to address structural inadequacy in their predictive validation framework \cite{oliver}.
It is useful to describe the ideas in \cite{oliver} by introducing an abstract prediction problem.

As mentioned, physical models are built upon reliable theory.
Let this theory be expressed mathematically as 
\begin{align}
	0 &= \mathcal{R}( c, \tau; s ). \label{eq:relTheory} 
\end{align}
This may be a system of equations; for example, \eqref{eq:relTheory} could represent conservation of mass, momentum, and energy.
Here, $c$ is the solution to \eqref{eq:relTheory}, $s$ is a collection of scenario parameters (geometry, initial condition, etc.), and $\tau$ represents one or more quantities that are necessary to solve \eqref{eq:relTheory}.
For instance, in fluid mechanics, $\tau$ could be the pressure field or the viscous stress tensor.

To calibrate and validate the physical model, it is necessary to have observations $d$. 
Assume there is an accurate observation operator $\mathcal{O}$ such that
\begin{align}
	d=\mathcal{O}(c, \tau; s ).
	\label{eq:obsOp}
\end{align}
Finally, to compute the quantity of interest $q$, there is assumed to be an operator $\mathcal{Q}$ such that
\begin{align}
	q=\mathcal{Q}(c,\tau; s).
	\label{eq:qoi}
\end{align}
Together, the prediction system is
\begin{align*}
	0 &= \mathcal{R}( c, \tau; s ) \\
	d &= \mathcal{O}(c, \tau; s ) \\
	q &= \mathcal{Q}(c, \tau; s ) . 
\end{align*}
If $\tau$ is unknown, this system is exact but unclosed.
To solve \eqref{eq:relTheory}, it is necessary to introduce a model for $\tau$:
\begin{align*}
	\tau \approx \tau_m( c; r, \theta ),
\end{align*}
where $r$ is a set of scenario parameters that may differ from $s$ and $\theta$ are a set of model parameters.
This model for $\tau$ could arise from a series of modeling assumptions, empirical correlations, or interpolation of data. 
Such models are only suitable in specific situations, such as when the simplifying assumptions used to derive them hold true. 
Outside such situations, the model's prediction of $\tau$ is suspect at best. 
It is the use of $\tau_m$ outside of its ``region of reliability'' that can introduce error into the composite model ($\mathcal{R}$ with $\tau_m$ substituted for $\tau$).

A common approach first introduced by Kennedy and O'Hagan to address this error appends a stochastic discrepancy term to calibration observables,
\begin{align*}
	0 &= \mathcal{R}( c, \tau_m; s ) \\
	d &= \mathcal{O}(c, \tau_m; s ) + \eps_m( s, \theta )\\
	q &= \mathcal{Q}(c, \tau_m; s ) ,
\end{align*}
where $\theta$ represent hyperparameters that are calibrated  \cite{kennedyOhagan}. 
This approach is insufficient for scenarios where there is no mapping between the observable quantities and the quantity of interest, because nothing can be gleaned about the effect of the model inadequacy on QoI predictions.

A more enabling approach is to instead enrich the embedded model $\tau_m$ with a stochastic error representation.
For instance, it could be enriched with an additive term:
\begin{align*}
	\tau \approx \tau_m + \eps_m.
\end{align*}
The full system of equations is thus
\begin{align*}
	0 &= \mathcal{R}( c, \tau_m + \eps_m; s ) \\
	d &= \mathcal{O}(c, \tau_m + \eps_m; s ) \\
	q &= \mathcal{Q}(c, \tau_m + \eps_m; s ) . 
\end{align*}
This propagates uncertainty to the QoI while simultaneously allowing calibration using available data.

\subsection{Application problem description} 
As a more concrete example, consider field-scale transport of a contaminant through a heterogeneous porous medium.
For the purposes of this research, the advection-diffusion equation (ADE) is considered a highly reliable model of this phenomenon.
The equations are
\begin{align}
    \diffp[]{c(\mathbf{x}, t)}{t} + \div{\ub(\mathbf{x})c(\mathbf{x},t)} = \nu_m \grad^2 c(\mathbf{x},t), \label{eq:detailedConsMass}\\
    \div{\ub} = 0, \label{eq:detailedContinuity}\\
    \ub(\mathbf{x}) = -\frac{\kappa(\mathbf{x})}{\phi \mu}\grad p(\mathbf{x}), 
\end{align}
where $\nu_m$ represents molecular diffusivity, $\kappa(\mathbf{x})$  permeability, $\phi$ porosity, and $\mu$ viscosity.
All constants are assumed known, and given initial and boundary conditions, the structure of the permeability field entirely determines the transport of the contaminant.
Generally, the entire permeability field is not known, but statistical information, such as the mean and correlation structure, may be available by analyzing cores taken from the medium.
The limited information about the underlying permeability field motivates taking a statistical mean of the equations.

The equations will also be averaged in the depthwise direction.  
This is for multiple reasons.
First, although the contaminant exists in 3D, observations of the contaminant are limited to a depthwise ($y$-direction) average due to mixing that occurs when drawing fluid from a well for measurement.
Second, the depthwise variation of the contaminant's concentration is not generally the relevant quantity of interest; of more concern is when or where in the streamwise direction the average depthwise concentration for a contaminant exceeds a tolerance.

To obtain a set of equations for the statistically and spatially averaged concentration, consider the probability space $(\Omega, \mathcal{F}, \mathbb{P})$ and a random field $f(\mathbf{x}, \omega)$.
For the random field $f(\mathbf{x},\omega)$ let
\begin{align*}
    \mean{ f(\mathbf{x}, \omega) } \equiv \int_0^{L_y} \int_{\Omega}f(\mathbf{x},\omega)\d \mathbb{P}(\omega) \d y.
\end{align*}
This leads to a Reynolds decomposition of the field, $f(\x, \omega) = \mean{f(\x,\omega)} + f'(\x,\omega)$.
Substituting this decomposition into the equations for $c, u$ and applying the averaging operator on \eqref{eq:detailedConsMass} and \eqref{eq:detailedContinuity} gives
\begin{align*}
    \diffp[]{\meanc}{t} + \mean{u}\cdot\grad\meanc + \grad\cdot\mean{u'c'} = \nu_m \grad^2 \mean{c}.
\end{align*}
This equation is exact but unclosed because of the fluctuation term $\mean{u'c'}$, which is called the dispersive flux.
Relating this back to the previous definitions, 
\begin{align*}
    \mathcal{R}(c,\tau) &= \diffp[]{\meanc}{t} + \mean{u}\cdot\grad\meanc + \grad\cdot\tau - \nu \grad^2 \mean{c} = 0, \\
    \tau &= \mean{u'c'}.
\end{align*}

A typical closure model for this term is a gradient-diffusion model, or Fick's first law \cite{bear2010modeling}
\begin{align*}
    \tau_m = - \nu' \grad\meanc. 
\end{align*}
Substituting this closure model into the reliable theory results in an ADE for the mean concentration:
\begin{equation}
    \begin{aligned}
        \diffp[]{\meanc}{t} + \mean{u}\cdot\grad \meanc = \nu \grad^2 \mean{c}, \\
        \nu = \nu_m + \nu'. 
    \end{aligned}
    \label{eq:1DADE}
\end{equation}

Dispersion is caused by local fluctuations from the mean fluid velocity that transport the contaminant to different locations in the medium at different speeds. 
The effect of this heterogeneous advection on the averaged concentration field is diffusive, which is why the Fickian model is often employed.
For field-scale transport through a heterogeneous porous medium, fluctuations from the mean can be quite large, and the evolution of the mean concentration will depend significantly on the fine-scale detail of the fluctuations.
In this case, the gradient-diffusion model for the dispersion will not reproduce the correct behavior.

It is well known that the Fickian model of dispersion is not appropriate for field-scale modeling of transport through heterogeneous media \cite{levy2003nonFickian, heterogeneousFlow, neuman2009perspective}.
Clearly, the gradient-diffusion model is the source of model inadequacy in \eqref{eq:1DADE}.
As a result, the error representation will appear as an enrichment of $\tau_m$:
  \begin{align*}
      \mean{u'c'}&= \tau_m + \epsilon_m(\mean{c}) \\
      &= \nu' \mean{u}\grad{c} + \epsilon_m(\mean{c}).
    \end{align*}
The reliable theory, with the augmented closure model, will be called the composite model:
\begin{equation}
    \begin{aligned}
        0 &= \mathcal{R}(c, \tau_m + \epsilon_m; s ) \\
        &= \diffp{\meanc}{t} + \meanu \cdot\grad\meanc - \nu_m \grad^2 \meanc + \div{ -\nu' \mean{u} \grad{\meanc} + \epsilon_m(\mean{c})}.
    \end{aligned}
    \label{eq:compositeModel}
\end{equation}

  \subsection{Model inadequacy representation}
This work focuses on developing the error representation, also called the inadequacy model. 
Specifically, the goal of this work is to develop an inadequacy representation that characterizes the uncertainty caused by using a gradient-diffusion model for transport through a heterogeneous medium.
The process of dispersion it is intended to represent depends significantly on fine-scale detail that is missing in the gradient-diffusion model, which is the cause of its inadequacy.

In the context of physics-based models, the inadequacy model must be constructed to respect physical knowledge while remaining flexible enough to be informed by available data.
The inadequacy model is made stochastic to represent uncertainty in model form that cannot be resolved deterministically because of missing physical information or missing physics in the model itself.
At a minimum the inadequacy representation must also be defined so that the composite model respects physical laws that the original model obeyed, such as conservation of mass.
Finally, the inadequacy representation should be state-dependent, so that as the state is extrapolated, so too is the representation.
Beyond these basic requirements, this research will explore methods of incorporating knowledge of the closure model's shortcomings.
Above all, it is crucial to keep in mind that the inadequate model will be employed in a predictive scenario, so physically-motivated constraints on the inadequacy representation are key to a successful extrapolation.

The Center for Predictive Engineering and Computational Science (PECOS) is pursuing multiple ways of formulating the stochastic error representation.
For example, one approach models the inadequacy as the solution to a stochastic PDE with terms for transport, diffusion, and so on, with random forcing.
An alternate approach models the inadequacy as a finite-dimensional stochastic operator \cite{morrison}. 
This work instead models the inadequacy as an infinite-dimensional linear operator. 
In the finite-dimensional case it is possible to constrain the operator by constraining its elements. 
In the infinite-dimensional case this is not an option, but it is possible to instead constrain its spectrum, thereby encoding information about the relevant physics and the inadequate embedded model.

All calibration data for this work was generated using higher-fidelity models that include a more accurate representation of the physics being left out of the low-fidelity model in question.
By using data generated by such models, all errors can be controlled, so that error caused by model inadequacy can be isolated.
In many physical problems of interest, such high-fidelity models exist but are too expensive to be used for engineering problems, so methodology using such high-fidelity models may be of use in other applications as well.

The remainder of the article proceeds as follows. 
Section 2 will outline the formulation and parametrization of the inadequacy model as a stochastic operator, a simple inverse problem to infer a deterministic operator, and a method of interrogating a high-fidelity model to learn about model error.
Section 3 will provide current results, 
and Section 4 will discuss conclusions and outline future work.

\section{Inadequacy Model Development}
In this section, the steps taken toward developing a framework for developing and calibrating a physically meaningful inadequacy operator representation are described.
For simplicity, the low-fidelity model used in work to this point has been the 1D ADE, modeling the streamwise transport of the statistically- and depthwise-averaged concentration field of a contaminant.
The formulation and parametrization of the inadequacy operator is defined in the next subsection.
The problem of inferring an infinite-dimensional deterministic operator has not been explored, so a simple problem is defined and explored in order to determine its feasibility and to uncover any inherent challenges.
Finally, a method of interrogating high-fidelity models is described, which will enable the formulation and calibration of the inadequacy operator.

\subsection{Inadequacy formulation and parametrization}
As mentioned in the discussion of the ADE, Fick's first law is an inadequate representation of dispersion through a heterogeneous porous medium. 
This is because the true dispersion depends on local fluctuations of the velocity and concentration fields.
The inadequacy operator must be stochastic to account for this missing information.

  It is important for extrapolation that the inadequacy representation depend on the state variable.
  This motivates representing $\epsilon_m(c)$ as a stochastic operator, denoted $\mathcal{L}$.
  The underlying physical process is linear, so the operator must also be linear to respect this constraint.
  Rearranging terms in the conservation law and including the boundary and initial conditions gives the enriched composite model:
  \begin{equation}
	\begin{aligned}
		\diffp{\meanc}{t} + \meanu \diffp{\meanc}{x} = \diffp{}{x}\left( \nu \diffp{}{x} + \mathcal{L} \right)\meanc,& \quad x\in (0,L_x) \\
		\meanc(0, t) = \meanc(L_x, t) &\\
		\meanc(x,0) = c_0(x).
	\end{aligned}
	\label{eq:linearModel}
\end{equation}

Under fairly general assumptions that the solution to the PDE lies in a separable space and the eigenfunctions of the linear operator form an orthonormal basis on that space, the operator can be fully characterized by its eigenvalues and eigenfunctions $\left\{ \lambda_k, \phi_k \right\}_{k=-\infty}^{\infty}$.
This work focuses on this parametrization of the inadequacy operator, which is leveraged to constrain and calibrate the operator.
Assuming this parametrization holds, one can express the action of $\mathcal{L}$ on $\meanc$ as
\begin{align*}
	\mathcal{L}\meanc &= \mathcal{L}\left( \sum_{k=-\infty}^{\infty}c_k \phi_k \right)\\
	&= \sum_{k=-\infty}^{\infty} c_k \lambda_k \phi_k,
\end{align*}
where $c_k$ are expansion coefficients.

The permeability field is assumed to be statistically homogeneous, so diffusion caused by flow through the medium should depend only on distance traveled.
This implies that the composite diffusion term, and thus $\mathcal{L}$, should be translation-invariant.
Translation-invariant linear operators have Fourier modes as eigenfunctions ($\phi_k(x) = \e{ik_x x}$), so the eigenvalues of $\mathcal{L}$ are the only unknowns in this problem.
  
The goal is to encode as much knowledge into the structure of $\mathcal{L}$ as possible.
This can be knowledge of relevant physics or of the model error. 
In certain cases, it is easier to constrain the behavior of the whole composite model rather than constraining $\mathcal{L}$ directly. 
For instance, consider the right-hand side of \eqref{eq:linearModel}. 
Define
\begin{align}
	\mathcal{D} :=	\diffp[]{}{x}\left( \nu \diffp[]{}{x} + \mathcal{L} \right)
	\label{eq:D}
\end{align}
and denote its eigenvalues $\mu_k$. 

While it is unclear exactly what the effect of $\mathcal{L}$ on $\meanc$ should be, it is known that the effect of $\mathcal{D}$ should be smoothing. 
This corresponds to the Fourier coefficients of the state variable, $\hat{c}_k(t)$ decaying with time.
Substituting the Fourier expansion of $\meanc$ into \eqref{eq:linearModel} gives the following expression for the Fourier coefficients: 
\begin{align*}
	\hat{c}_k(t) = \hat{c}_k(0)\e{\left[\mu_k - \meanu \itpk\right]t}, \quad k \in \Z,
\end{align*}
where $\hat{c}_k(0)$ are the Fourier coefficients for the initial condition.
Their magnitudes are
\begin{align*}
	\abs{ \hat{c}_k(t) } & = \abs{ \hat{c}_k(0)\e{[\mu_k - \meanu \itpk]t} }\\
	&= \abs{\hat{c}_k(0)}\abs{\e{\Re[\mu_k]t} }\abs{\e{i\left(\Im[\mu_k] - \meanu \tpk\right) t}},
\end{align*}
so requiring that $\Re[\mu_k] \leq 0$ guarantees that the coefficients will decay with time.
With this in mind, calibration will be performed on $\mu_k$ and $\lambda_k$ will be recovered via the relation
\begin{align*}
	\mu_k = \nu \itpk^2 + \lambda_k \itpk.
\end{align*}

The requirement that the eigenvalues of $\mathcal{D}$ lie in the left-half plane naturally suggests a further parametrization of $\mu_k$ via their polar form:
\begin{align*}
	\mu_k = r_k e^{i\theta_k},
\end{align*}
where bounds are placed on the arguments of the eigenvalues, $\theta_k\in[\pi/2, 3\pi/2]$ and $r_k$ must be positive.

In principle this is an infinite-dimensional sampling problem, but in reality it becomes finite-dimensional when the Fourier expansion is truncated in order to compute a numerical solution.
The state variable $c$ is real, so the Fourier coefficients are conjugate symmetric.
Given that there are two parameters, $r_k$ and $\theta_k$ for every wave number, the problem size is thus equal to the number of Fourier modes left in the Fourier expansion after truncation.

To ensure the operator is stochastic, the eigenvalues must be defined so that they are stochastic.
There are many ways to achieve this, but in this case the eigenvalues will likely be modeled as solutions to stochastic ODEs.
  
\subsection{Inferring an unknown deterministic operator}
Before jumping in and attempting to calibrate the eigenvalues of a stochastic operator, preliminary research has focused on inferring the eigenvalues of a deterministic operator. 
The first inverse problem was defined so that an analytical expression for the eigenvalues was available. 
This helped to determine if inference was successful and to identify challenges inherent to inferring an operator.

The unknown operator is $\mathcal{D}$, defined in \eqref{eq:D}. 
Its eigenvalues $\mu_k$ are parametrized as described in the previous subsection but are assumed deterministic for this exercise.
The generalized ADE is
\begin{equation}
\begin{aligned}
	\diffp[]{\meanc}{t} + \meanu \diffp[]{\meanc}{x} = \mathcal{D}\meanc,& \quad x\in(0,L_x) \\
		\meanc(0, t) = \meanc(L_x, t), &\\
		\meanc(x,0) = c_0(x). &
\end{aligned}
	\label{eq:generalizedADE}
\end{equation}

  \subsubsection{1D data model}

An alternative approach to modeling contaminant transport that has gained traction makes use of fractional differential equations. 
Such models can be seen as limiting forms to solutions of the more general and promising approach continuous time random walks to models of non-Fickian transport \cite{berkowitz2006ctrw}.
Specifically, a fractional differential model of the dispersive flux results in the fractional advection-diffusion equation (FRADE)
\begin{equation}
	\begin{aligned}
		\diffp{\meanc}{t} + \meanu \diffp{\meanc}{x} = \nu \diffp[\alpha]{\meanc}{x},& \quad x\in (0,1), \quad \alpha \in [1,2] \\
		\meanc(0, t) = \meanc(L, t), &\\
		\meanc(x,0) = \exp( -(x-x_0 )^2 / l^2 ). &
	\end{aligned}
	\label{eq:FRADE}
\end{equation}
An example of the time evolution of the concentration field using this model is shown in Figure \ref{fig:FRADEevolution}.
The eigenvalues of $\mathcal{D}$ can be determined analytically in this case by setting the right-hand sides of \eqref{eq:generalizedADE} and \eqref{eq:FRADE} equal to each other:
\begin{align*}
	\mu_k = \nu \itpk^{\alpha}.
\end{align*}
\noindent%
\begin{minipage}{\linewidth}
	\makebox[\linewidth]{
	  \includegraphics{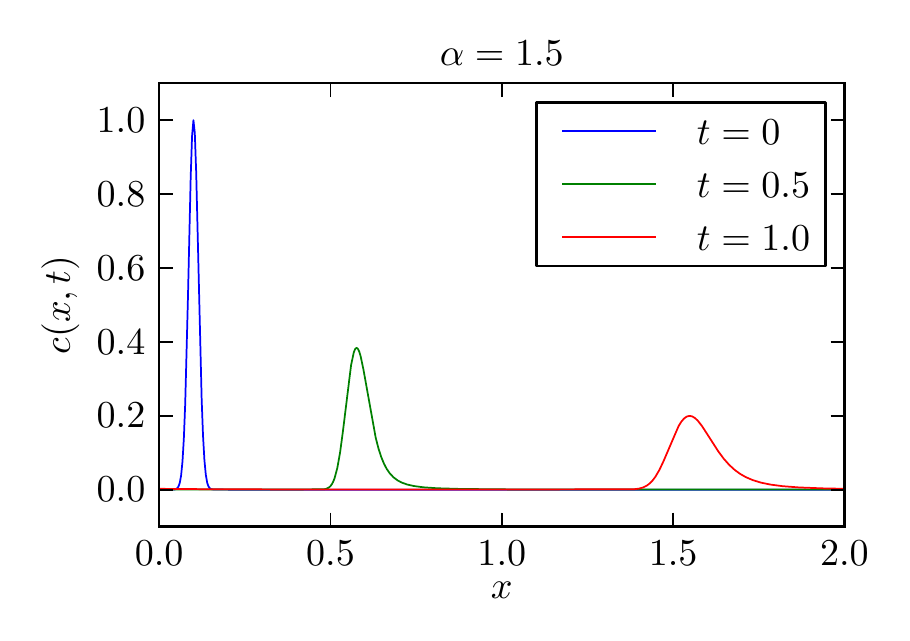}
	}\vspace{-0.5cm}
    \captionof{figure}{The FRADE model reflects the asymmetric diffusion of the concentration field.}
	  \label{fig:FRADEevolution}
	  \end{minipage}

The eigenvalues were calibrated with and without an i.i.d.~measurement error assumption via a Bayesian and a deterministic optimization, respectively.
Having access to the true set of eigenvalues helped to determine if the inference was successful and helped to identify challenges in the calibration process. 
These challenges will be discussed in more detail along with the results in \ref{fradeResults}.

  \subsubsection{2D data model}
\begin{figure}[h]
	  \centering
	  \includegraphics{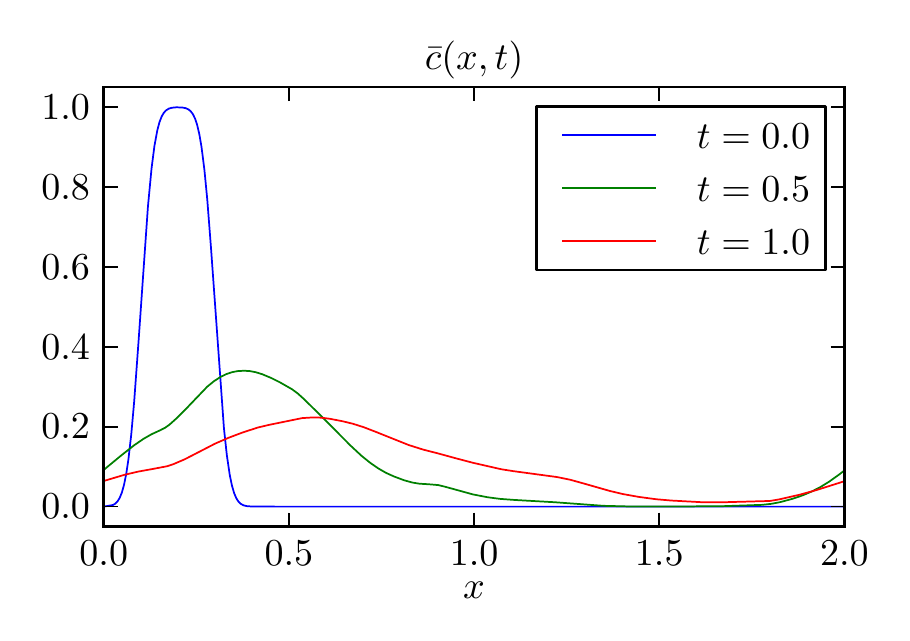}
	  \caption{The diffused concentration field is not as smooth as predicted by FRADE or standard ADE models.}
	  \label{fig:adeSolns}
  \end{figure}

  The second calibration scenario explores the optimal linear shift-invariant operator to represent diffusion for \eqref{eq:generalizedADE}, given data from a single permeability field (instead of an ensemble average).
 In this case, there is no set of eigenvalues for which the solution of \eqref{eq:generalizedADE} will exactly match with the data generated from a 2D model. 
The 2D ADE with incompressible Darcy flow was used to generate calibration data:
\begin{equation}
\begin{aligned}
    &	\diffp[]{c(\x,t)}{t} + \ub(\x) \cdot \grad c(\x,t) = \nu \Delta c(\x,t), \quad \x \in [0,L_x]\times[0,L_y]\\
	& \ub(\x) = - \frac{\kappa(\x)}{\phi \mu} \grad p(x),  \quad
    \div{\ub(\x)} = 0, \\
&	c(0,y, t) = c(L_x, y, t), \quad 
\at{ \diffp[]{c}{y} }{y=0} = \at{ \diffp[]{c}{y} }{y=L_y} = 0 \\
&	c(x,y,0) = c_0(x). \\
\end{aligned}
	\label{eq:2DperiodicADE}
\end{equation}
The permeability field is randomly generated so that it is periodic in the streamwise direction and aperiodic in the vertical direction.
The 2D solution of \eqref{eq:2DperiodicADE} is then averaged in the depth-wise direction:
\begin{align*}
	\bar{c}(x, t) = \frac{1}{L_y}\int_0^{L_y} c(x,y, t) dy.
\end{align*}
See Figure \ref{fig:adeSolns} for examples of the upscaled concentration's evolution through this model.

  \subsection{Hessian-Informed Calibration}
  The generalized ADE \eqref{eq:generalizedADE} is nonlinear in the calibration parameters, which can cause simple algorithms to struggle. 
  The analytical expression for the Hessian of the data misfit function is available and has been leveraged for both the deterministic calibration and the Bayesian calibration of the operator.
  A Newton optimization is used for the deterministic calibration or to seed at the MAP point for the Bayesian calibration.
  Derivative information is provided to the manifold MALA (MMALA) algorithm implemented in the MIT UQ library; this is equivalent to the Stochastic Newton MCMC algorithm, if the analytical Hessian is used \cite{MMALA, MUQ, stochasticNewton}. 
  Let $\rb:=[r_1, r_2, \cdots, r_{n_k} ]$ and $\thetab := [ \theta_1, \theta_2, \cdots, \theta_{n_k}]$ be vectors of the radii and arguments of the eigenvalues $\mu_k$. 
  Furthermore let $\Thetab := [ \;\rb,\; \thetab\; ]\in\R^{2n_k}$ be the full vector of parameters on which the state variable $\meanc$ depends parametrically. 
  The vector of discrete samples of the state variable is $\cb := [ \;\meanc(x_1, t_1), \meanc(x_2, t_2), \cdots, \meanc(x_{\nobs}, t_{\nobs})\;] $.
  The data vector $\db$ is equivalently defined.
  The misfit function is a standard squared $l^2$ norm:
  \begin{align*}
	  J( \Thetab ) := \frac{1}{2} \norm{ \mathbf{c}(\Thetab) - \db }_{2}^2.
  \end{align*}
 The Hessian of the misfit function can thus be written in block form as 
 \begin{align*}
     H = \mtx{c|c}{ H_{rr} & H_{r\theta}
     \\\hline \rule{0pt}{\normalbaselineskip} H_{r \theta}^T & H_{\theta \theta} }.
 \end{align*}
 Each of these sub-Hessians can further be written as the sum of a misfit term and a Gauss-Newton Hessian term, e.g.
 \begin{align*}
	 H_{rr} &= H_{rr}^{misfit} + H_{rr}^{GN} 
	 = (\nabla_{rr}  \cb )^{T} (\cb - \db ) + ( \nabla_r \cb )^{T} (\nabla_r \cb ).
 \end{align*}
The Jacobian of the parameter-to-observable map is $\grad_{\Theta} \cb = [\;\grad_r \cb,\; \grad_\theta \cb\;]\in \R^{n_{obs} \times 2n_k}$, where
\begin{align*}
    \mtx{c}{
        (\grad_r \cb)_{jk} \\ 
(\grad_{\theta} \cb )_{jk}
}^T
& = \mtx{c}{
    \diffp[]{c_j}{r_k} \vspace{.5em}\\
\diffp[]{c_j}{\theta_k}
}^T
=  2\Re\left( \mtx{c}{
\e{i\theta_k}t_j  \\
i t_j \mu_k
}^T
\hat{c}_k(t_j) \phi_k(x_j) \right).
\end{align*}
The second derivative of the state variable with respect to the parameters at a given observation $(x_j, t_j)$ is a block matrix like $H$, each submatrix a diagonal matrix, where the entries are

\begin{align*}
    \mtx{c}{
        (\grad_{rr}\cb)_{kk} \\
        (\grad_{\theta\theta}\cb)_{kk} \\
        (\grad_{\theta r}\cb)_{kk}
}
&= 
2\Re \left(
    \mtx{c}{
        (\e{i\theta_k} t_j )^2 \\
        -t_j \mu_k(1+ r_k t_j) \\
it_j \e{i\theta_k}\left( 1+ t_j \mu_k \right) 
} 
\hat{c}_k(t_j) \phi_k(x_j) 
\right)
.
\end{align*}

  \subsection{Dimension reduction}
  As was stated previously, the infinite-dimensional problem is reduced to a finite-dimensional problem with the truncation of the Fourier expansion of $\meanc$. 
  Furthermore, as might be expected, the smoothing nature of the governing PDE causes the data misfit function to be sensitive to only a small subset of the parameters.
  With this in mind, only the subset of parameters for which the data misfit function is deemed sensitive are calibrated.

  A rudimentary sensitivity analysis is performed separately for the radii and the arguments because the radii scale with wave number while the arguments do not.
Let $\Thetab_0$ be the starting point of the calibration, and define $(\rb^{sens})_{i} := ( H_{rr}(\Thetab_0) )_{ii} $ and $(\thetab^{sens})_{i} := ( H_{\theta\theta}(\Thetab_0))_{ii} $.
The idea is that the Hessian of the data misfit function is largest for those variables to which it is the most sensitive.
  
A cutoff index is determined for the radius ($k_r$) and the argument ($k_{\theta}$) by finding the highest wavenumber such that
  \begin{align*}
	  \rb^{sens}_{k_r} \geq \gamma_{tol} \max ( \rb^{sens} ), \quad\quad \thetab^{sens}_{k_\theta} \geq \gamma_{tol} \max(\thetab^{sens}),
  \end{align*}
  where $\gamma_{tol}$ is a relative tolerance that is usually set to $10^{-2}$, and taking the maximum of $k_r$ and $k_\theta$.

  This method has reduced the dimension of the parameter space significantly in both calibrations, from 500-1024 parameters to less than 50.
  The optimization did not converge without this reduction.
  More sophisticated techniques for dimension reduction were considered, but this method was attempted first because of its relative simplicity, and it proved sufficient for the purposes of the current study.

  \subsection{Rescaling calibration parameters}
  Parametrizing the eigenvalues using their radii and arguments leads to a drastic variation in scales across the parameter space.
  The radii scale roughly with the wavenumber, while the arguments are independent of wavenumber.
  Rescaled parameters were introduced to reduce this difference, following a similar methodology to defining non-dimensional variable for a physical problem:
  \begin{align*}
	  r^*_k &= \frac{r_k - r_0}{\Delta r_k }, \quad  \quad
	  \theta^*_k = \frac{\theta_k - \theta_0}{\Delta \theta_k}.
  \end{align*}
  
  Determining appropriate values for $\theta_0$ and $\Delta \theta_k$ was straightforward because $\theta_k$ were defined to fall in the range $[\pi/2, 3\pi/2]$.
  In this case, $\theta_0 = \pi/2$ and $\Delta \theta_k = \pi$. 
  Unlike the argument, the radius is not formally bounded by any modeling assumptions, so the radii of the advection operator $\tpk$ and the diffusion operator $\tpk^2$ were used to determine values for $r_0$ and $\Delta r_k$. 
  The radius of the general diffusion operator is most likely greater than that of the radius for pure advection, so $r_0 = \tpk$. 
 The scale was taken to be $\Delta r_k = \tpk^2 - \tpk$, although it is possible that the radius could be larger than that of the diffusion operator.
 Then the final expressions for the rescaled variables are 
 \begin{equation}
 \begin{aligned}
	 r^*_k &= \frac{r_k - \tpk}{\tpk^2 - \tpk}, \quad \quad
	 \theta^*_k = \frac{\theta_k - \pi/2}{\pi}.
 \end{aligned}
	 \label{eq:rescaledParams}
 \end{equation}
All results will be in the context of these rescaled parameters.

\subsection{Interrogating the high-fidelity model}\label{interrogation}
Access to a high-fidelity model enables a much more in-depth inspection of the inadequacy of the low-fidelity model.
By evolving the same initial condition through both models, the model error for the given IC is completely isolated and its evolution in turn can be inspected for inadequacy model development.
By choosing an eigenfunction of the low-fidelity model as an initial condition, even more information can be extracted.
Assumptions for the low-fidelity model correspond to verifiable mathematical behavior for the evolution of an eigenfunction.
The propagation of the eigenfunction through the high-fidelity model can be compared against these predictions to test the validity of the modeling assumptions or to systematically isolate the effects of an assumption being violated.

This approach was used to test assumptions of time-independence and shift-invariance of the diffusion operator.
Substituting the IC $c_0(x) = e^{ik'_x x}$ into \eqref{eq:generalizedADE} gives
         \begin{align*}
         c(x,t) &= \sum_{k=-\infty}^{\infty}\delta( k - k')\e{[\mu_k - \mean{u}(ik_x)]t}e^{ik_x x    } \\
         &= \e{[\mu_k'- \mean{u}(ik_x')]t}e^{ik_x' x}.
     \end{align*}
     \begin{align}
         \hat{c}_k(t) = \left\{ 
             \begin{array}{ll}
                 \e{[\mu_k' - \meanu (ik'_x)]t }, & k = k',\\
                 0 & k \not=k'
             \end{array}
             \right.
             \label{eq:fmodeICcoeff}
     \end{align}
Recall that the operator was assumed shift invariant because it was acting on mean quantities, and the mean of the permeability field was assumed homogeneous.
This assumption was violated by using a single permeability field for the high-fidelity model; although the mean is homogeneous, individual fields are heterogeneous.
This assumption was known to be violated a priori, so it was expected that other Fourier modes would be excited in the evolution of a single mode through the high-fidelity model.

On the other hand, the assumption of time independence for the diffusion operator was tested for validity, without knowing if it would hold beforehand.
Time-independence of the diffusion operator implies constant eigenvalues and a specific time evolution of the Fourier coefficient for a single Fourier mode initial condition.
Differentiating \eqref{eq:fmodeICcoeff} gives
\begin{align*}
	\diffp[]{\hat{c_k}}{t}  = \left( \mu_k - \meanu\itpkLx \right)\hat{c}_k(t).
\end{align*}
If $\mu_k$ are constants, one would expect the log derivative to be constant with time, since
\begin{align}
    \hat{c}_k(t)^{-1}\diffp[]{\hat{c}_k}{t}  = \mu_k - \meanu\itpkLx .
    \label{eq:logDeriv}
\end{align}
The log derivative of the Fourier coefficients for different wavenumber was computed as a function of time to test this assumption.
The results of this initial interrogation are discussed in \ref{interrogationResults}.

\section{Results and discussion} \label{results}
  \subsection{1D data model} \label{fradeResults}
  The deterministic operator $\mathcal{D}$ was calibrated using data with and without measurement error. 
  Unsurprisingly, the success of the inference depended highly on the amount and type of data available.
Inference was most successful using spatial data at a fixed time instead of time series data at a fixed location. 

\noindent%
\begin{minipage}{\linewidth}
	\makebox[\linewidth]{
	  \includegraphics{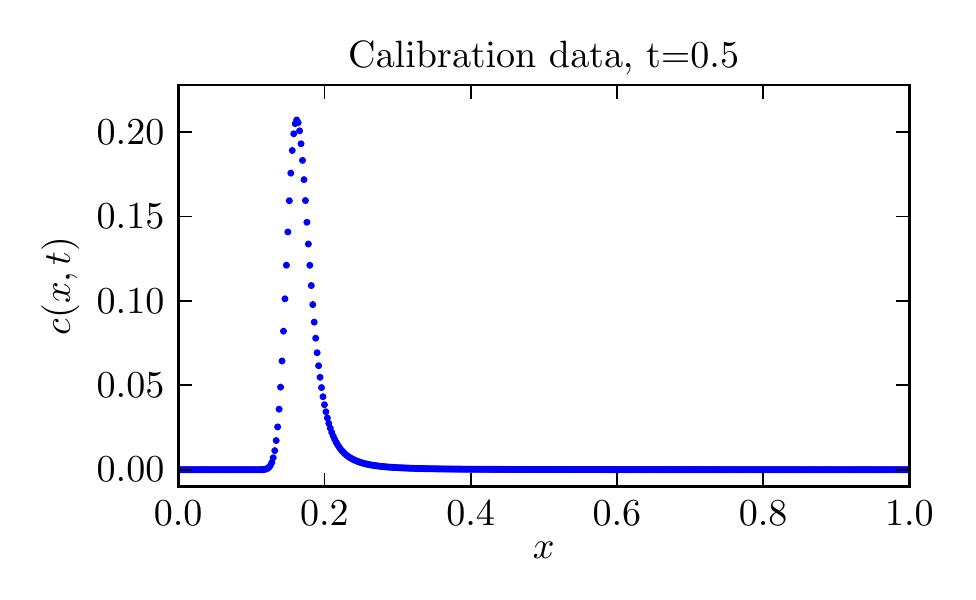}
	}\vspace{-0.5cm}
	  \end{minipage}

  \subsubsection{No observation error}
 In the best-case scenario with no measurement error and abundant calibration data, only the eigenvalues to which the objective function was sensitive could be informed.
 The smoothing nature of the diffusion operator means that the number of Fourier modes that are excited decreases as the solution evolves in time, so the later that data was taken, the fewer eigenvalues were informed.
  Only a subset of the eigenvalues were calibrated, selected using the sensitivity analysis technique described previously, as shown in Figure \ref{fig:deterministicFradeMus}.
  The rest of the eigenvalues were left at the values of the initial guess prior to optimization. 
  The initial guess for the results presented here was the eigenvalues of the second derivative operator.

\noindent%
\begin{minipage}{\linewidth}
	\makebox[\linewidth]{
	  \includegraphics{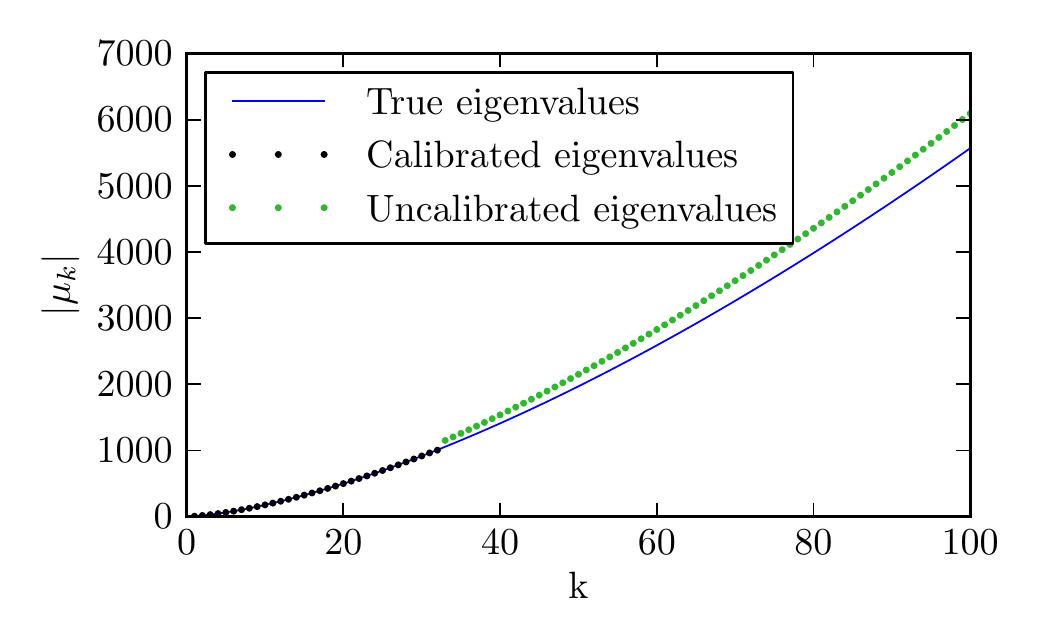}
	}\vspace{-0.5cm}
    \captionof{figure}{The first 32 eigenvalues were deemed significant during sensitivity analysis and calibrated in this case.}
	  \label{fig:deterministicFradeMus}
	  \end{minipage}
	  \linebreak\linebreak
  
  The eigenvalues that were informed by the data converged to the true values of the eigenvalues used to generate the calibration data.
  The resulting operator was used to evolve the initial condition to time $t=0.5$ and later times, and the evolution replicates that of the FRADE, as shown in Figure \ref{fig:deterministicFradeSolns}.
  This indicates that all the eigenvalues significant to the solution were inferred.
This test provides confidence that inference of data-informed eigenvalues is successful when there is abundant synthetic data available. 

\noindent%
\begin{minipage}{\linewidth}
	\makebox[\linewidth]{
	  \includegraphics{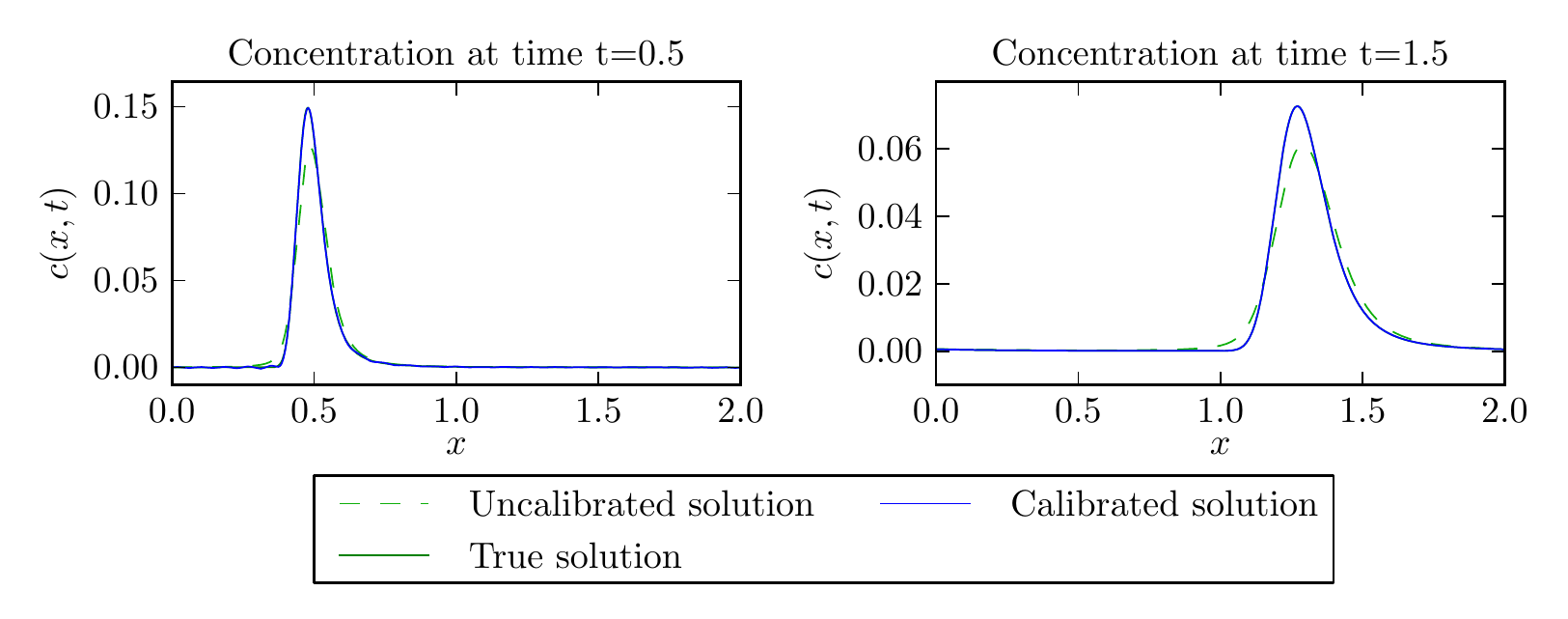}
	}\vspace{-0.5cm}
\captionof{figure}{The operator was calibrated with data from time $t=0.5$, then the calibrated operator was used to evolve the initial condition to time $t=1.5$.}
	  \label{fig:deterministicFradeSolns}
	  \end{minipage}
 
  \subsubsection{With observation error}
   The inclusion of observation error revealed a weakness in the current operator's formulation: the operator is not currently constrained to preserve positivity.
   A time series of the FRADE solution at one location contaminated by i.i.d.~Gaussian noise was used for calibration data.
   The eigenvalues were inferred via a Bayesian calibration, optimized over the full parameter space for the MAP point beforehand.
   A subset of the parameters was calibrated based on sensitivity analysis as previously described.
   Uniform $[0,1]$ priors were specified for all of the rescaled parameters.
   The MAP point of the eigenvalues' posterior distributions was used to evolve the concentration at the same location, and the calibrated solution coincides well with the time-series data; see Figure \ref{fig:inconsistentTimeSeries}.

   \noindent%
\begin{minipage}{\linewidth}
	\makebox[\linewidth]{
	   \includegraphics{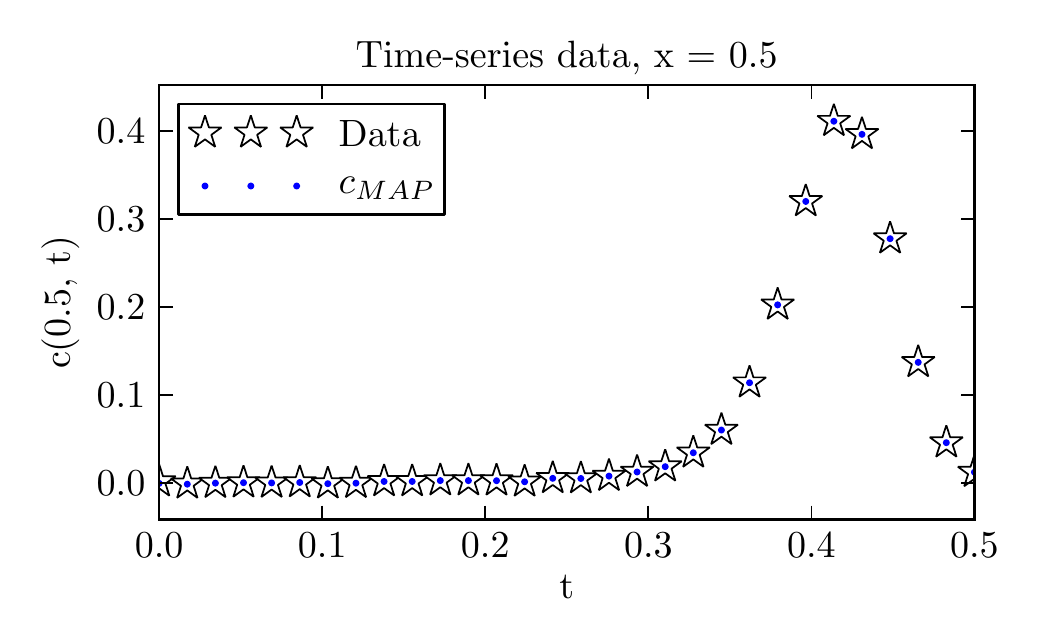}
   }\vspace{-.6cm}
    \captionof{figure}{MAP point of the eigenvalues' posterior distribution propagated to the time-series samples of the concentration field.}
	   \label{fig:inconsistentTimeSeries}
	  \end{minipage}
      \vspace{.2cm}

   The histograms of the first radii and arguments of the first few eigenvalues, however, show in Figure \ref{fig:inconsistentHistograms} that there is an issue. 
The posteriors are inconsistent with or assign low probability to the true values of the eigenvalues, although the concentration at the MAP point reproduces the time-series data well.

\noindent%
\begin{minipage}{\linewidth}
	\makebox[\linewidth]{
		\includegraphics{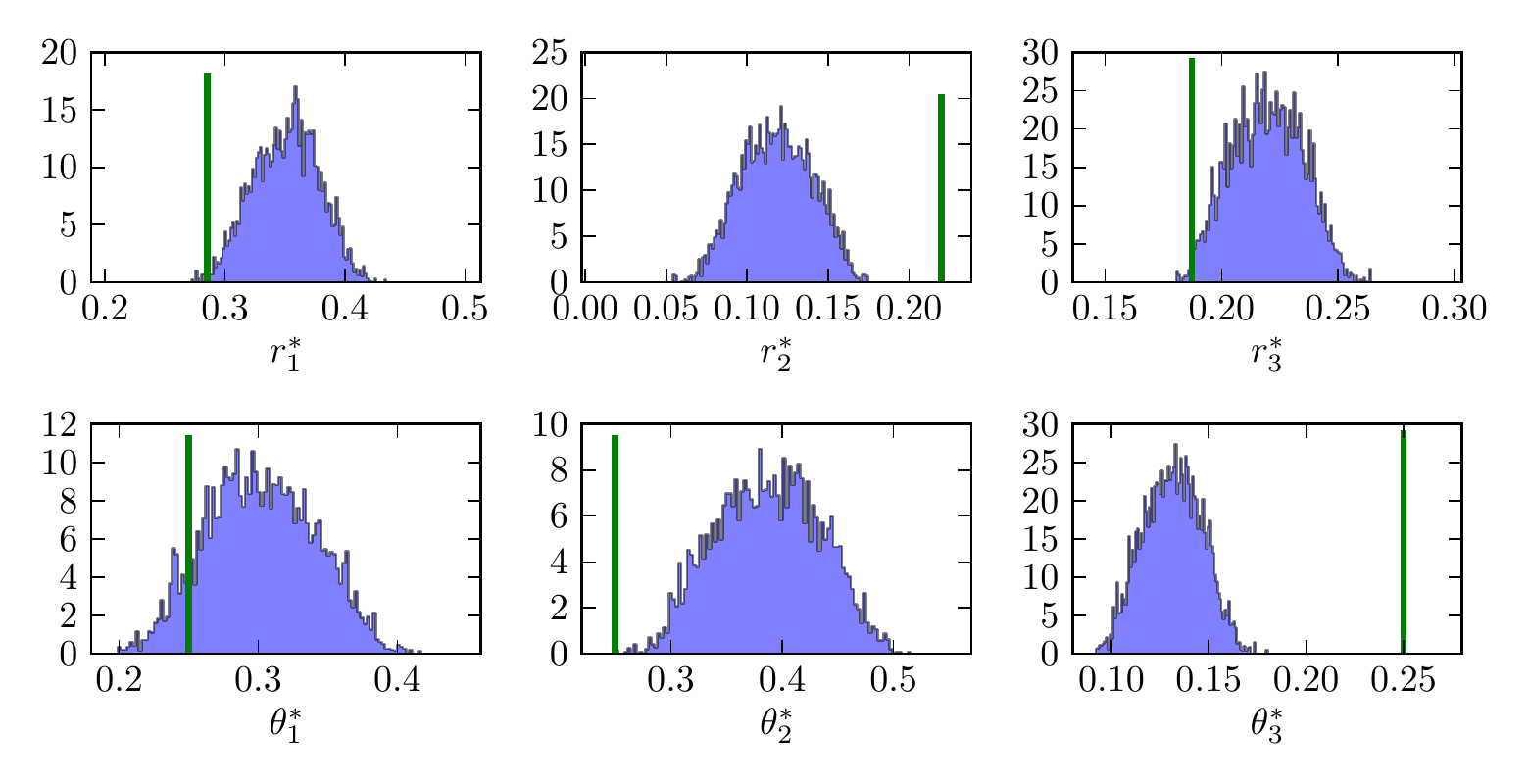}
	}\vspace{-0.5cm}
	   \captionof{figure}{Histograms for the radius and argument parameters corresponding to the first few eigenvalues after calibration with time-series data from $x=0.5$.}
	   \label{fig:inconsistentHistograms}
	  \end{minipage}
      \vspace{.2cm}

	  Using the MAP point the posterior to evolve the concentration field to $t=0.5$ in Figure \ref{fig:inconsistentSpatialSolution} shows what went wrong with the calibration. 
The time-series data was not enough to constrain the calibrated operator to physically meaningful solutions, allowing the calibration to fit to noise.
The evolution of the concentration field using the MAP point eigenvalues led to oscillations and negative concentrations, which is physically impossible.
The current formulation allows for regions of the parameter space to produce nonphysical evolutions.

\noindent%
\begin{minipage}{\linewidth}
	\makebox[\linewidth]{
	   \includegraphics{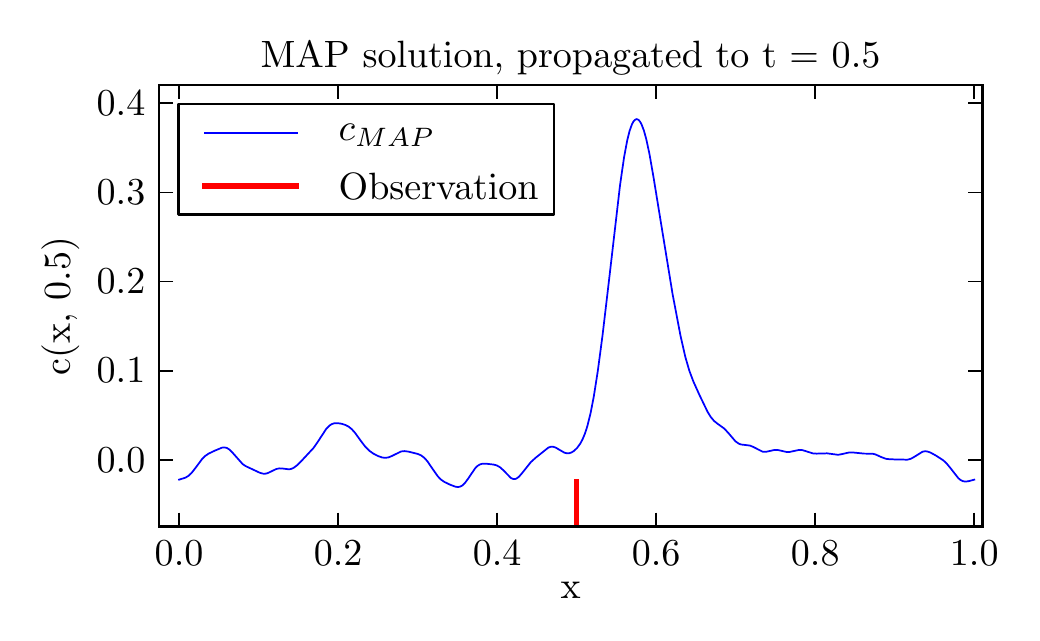}
	}\vspace{-0.5cm}
    \captionof{figure}{MAP point of the eigenvalues' posterior distribution propagated to the concentration field at time $t=0.5$.}
	   \label{fig:inconsistentSpatialSolution}
	  \end{minipage}
      \vspace{.2cm}

   On the other hand, using spatial data over the whole domain, as shown in Figure \ref{fig:consistentSolution}, does not allow these oscillations, constraining the eigenvalues to physical evolutions.
   Using the spatial data, the posterior distributions are consistent with the true eigenvalues (see Figure \ref{fig:consistentHistograms}).

	  \noindent%
\begin{minipage}{\linewidth}
	\makebox[\linewidth]{
	   \includegraphics{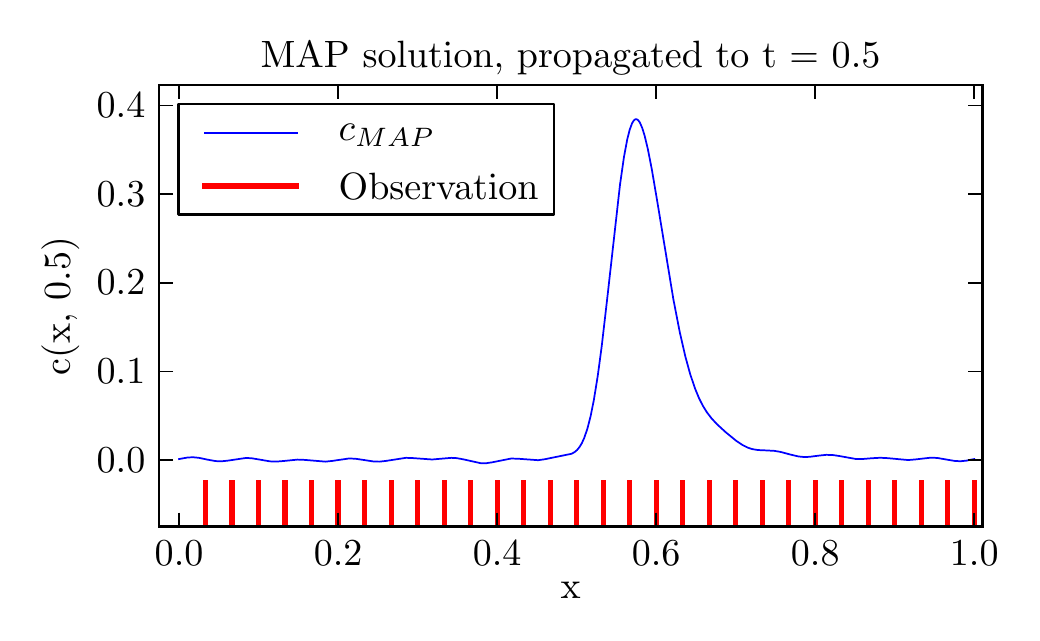}\vspace{-.2cm}
   }\vspace{-0.3cm}
   \captionof{figure}{The solution was observed at 30 spatial locations across the domain, which severely limited possible oscillations.}
	   \label{fig:consistentSolution}
   \end{minipage}
   \linebreak
   \linebreak

   The result of this exploration shows that, with limited spatial data, constraining the operator to preserve positivity is necessary for accurate results.
   However, with sufficient spatial data, this is not necessary for a successful calibration.
   Conditions on the eigenvalues of the operator that guarantee it preserves positivity will be complex and costly to impose. 
   Because higher-fidelity models will be used to generate data for this work, there is more than enough spatial data available for the calibration.
   With this in mind, we decided to side-step imposing such constraints, but with sparser or only time-series data, enforcing this constraint would be crucial.
	  \noindent%
\begin{minipage}{\linewidth}
	\makebox[\linewidth]{
	   \includegraphics{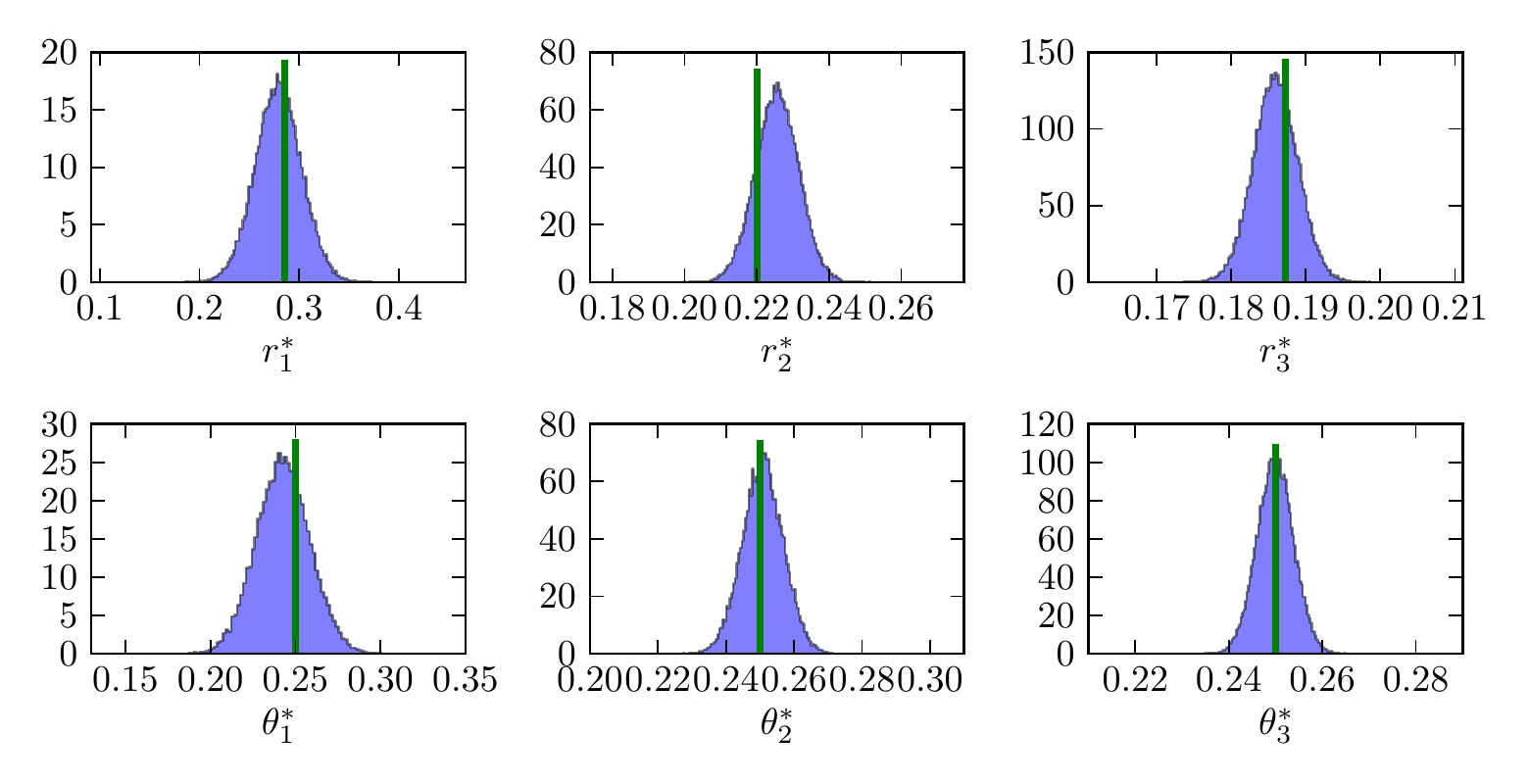}\vspace{-.2cm}
   }\vspace{-0.3cm}
   \captionof{figure}{Inference was much more successful and resulted with posteriors consistent with the truth when using more spatial observations.}
	   \label{fig:consistentHistograms}
   \end{minipage}
   \linebreak
   \linebreak

  \subsection{2D data model}
  After the successful calibration of the operator using synthetic spatial data, the same technique was applied to explore what the optimal deterministic operator would be using upscaled data from the 2D ADE.
  This data was generated by computing the 2D concentration field for a single permeability field and upscaling in the depthwise ($y$) direction.
  The resulting calibrated operator differs significantly from a second derivative operator or a fractional derivative (see Figure \ref{fig:2dparams}).

	  \noindent%
\begin{minipage}{\linewidth}
	\makebox[\linewidth]{
	\includegraphics{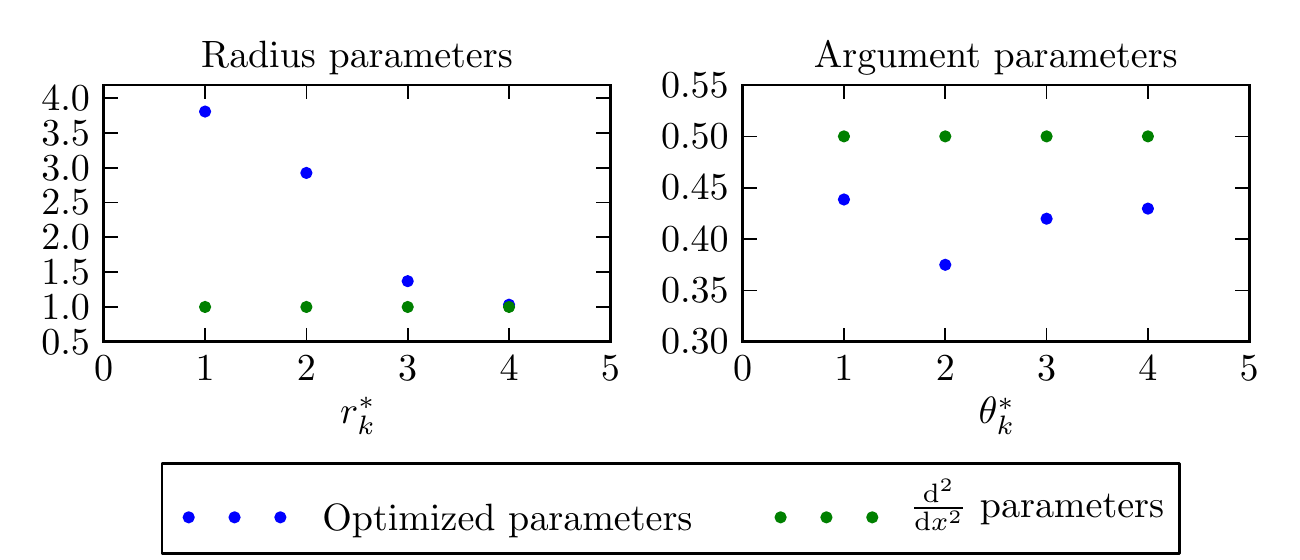}
}\vspace{-.2cm}
\captionof{figure}{Parameters after a deterministic calibration using samples of $\bar{c}(x, 1)$ as data.}
	  \label{fig:2dparams}
  \end{minipage}
  \vspace{.2cm}

  The rescaled radii are much larger than the second derivative operator and other fractional operators, and the rescaled radii decay rapidly compared to fractional operators. 
  The reason for this dramatic difference is yet unclear and inspired an interrogation of the high-fidelity model as described in \ref{interrogation}.
The calibrated operator was used to evolve the 1D ADE to different times and compared with data.
As shown in Figure \ref{fig:propagatedSolns}, the operator can be calibrated to fit the data fairly well at one time, but it fails to extrapolate in time.

	  \noindent%
\begin{minipage}{\linewidth}
	\makebox[\linewidth]{
		\includegraphics{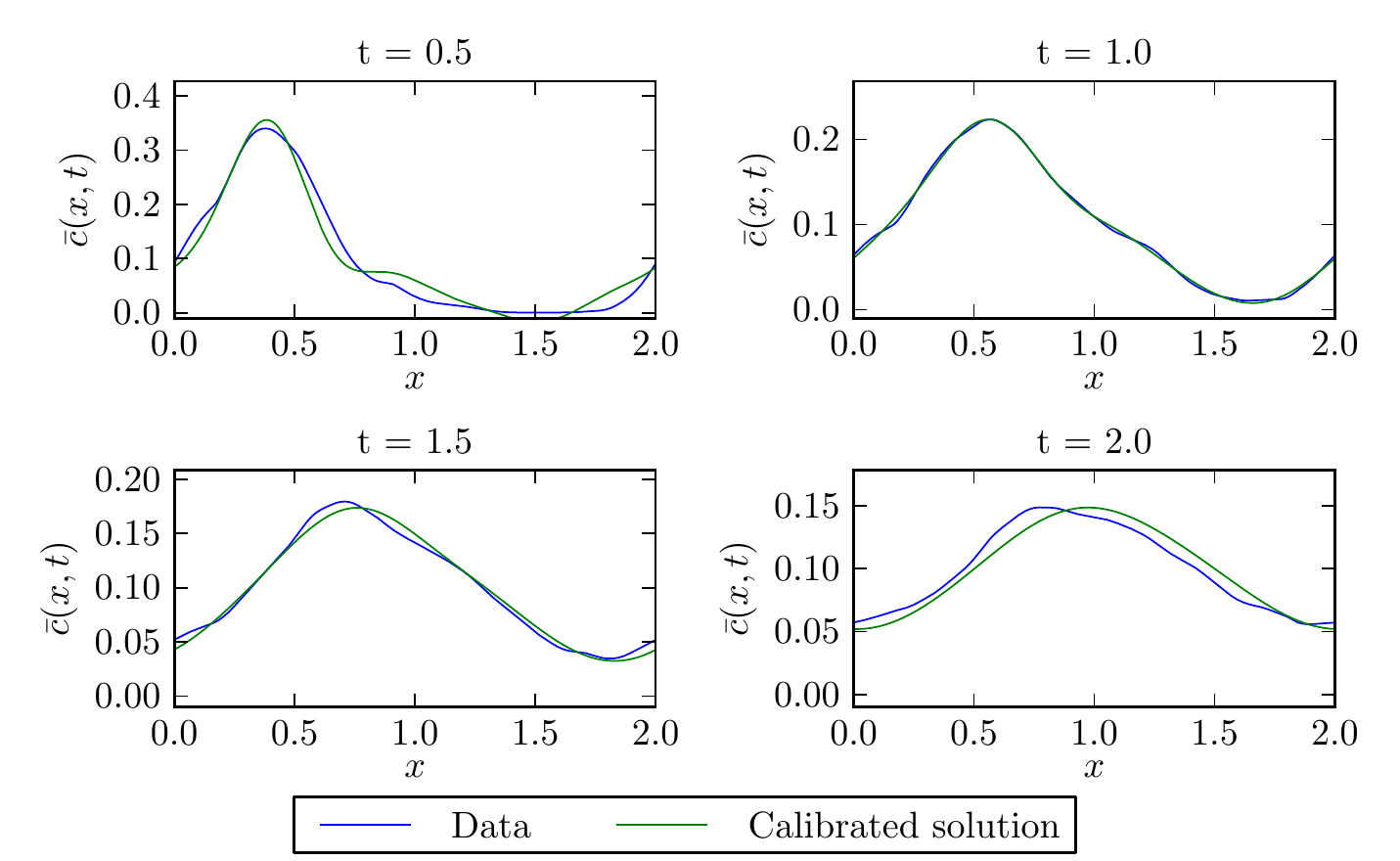}
    }\vspace{-.1cm}
	\captionof{figure}{The operator was calibrated using the data from $t=1$. The calibrated operator was then used to evolve the concentration field to $t=0.5, 1, 1.5$ and $2$.}
	  \label{fig:propagatedSolns}
  \end{minipage}
  \vspace{.2cm}

The discrepancy between the evolutions can be partially explained by the fact that the data was generated using only one permeability field instead of an ensemble average.
The shift-invariance assumption on the diffusion operator is only valid if the underlying field is homogeneous, which only holds for the permeability field in expectation.
There is additional information lost about fluctuations in the depthwise direction in the 1D model that could affect the model's ability to reproduce the evolution from the 2D model.

An ensemble average over permeability fields will be used as calibration data for the deterministic operator in the future.
The results of that calibration will be compared to those presented here, using a single permeability field to generate the data.
It will be particularly interesting to see if this unexpected structure of the eigenvalues persists with the ensemble average.

  \subsection{2D ADE Interrogation}\label{interrogationResults}
 
  As described in \ref{interrogation}, individual Fourier modes were propagated through \eqref{eq:2DperiodicADE} for a single permeability field and the log derivative of their Fourier coefficients computed.
  If the assumption of time-independence held for the diffusion operator, the log derivative would be constant with respect to time.
  However, as shown in Figure \ref{fig:log_coeffs}, the log derivative varies a great deal.

  Fourier coefficients for other modes were excited when propagating a single Fourier mode through the high-fidelity model, showing that the shift-invariance assumption is violated for transport through a single, heterogeneous permeability field.
Some of the variation in the log derivative in the coefficients may attributed to the excitation of the other modes.
The study will be repeated, computing an ensemble average over permeability fields for each Fourier mode. 
This will isolate any errors caused by upscaling in the depthwise direction.

	  \noindent%
\begin{minipage}{\linewidth}
	\makebox[\linewidth]{
		\includegraphics{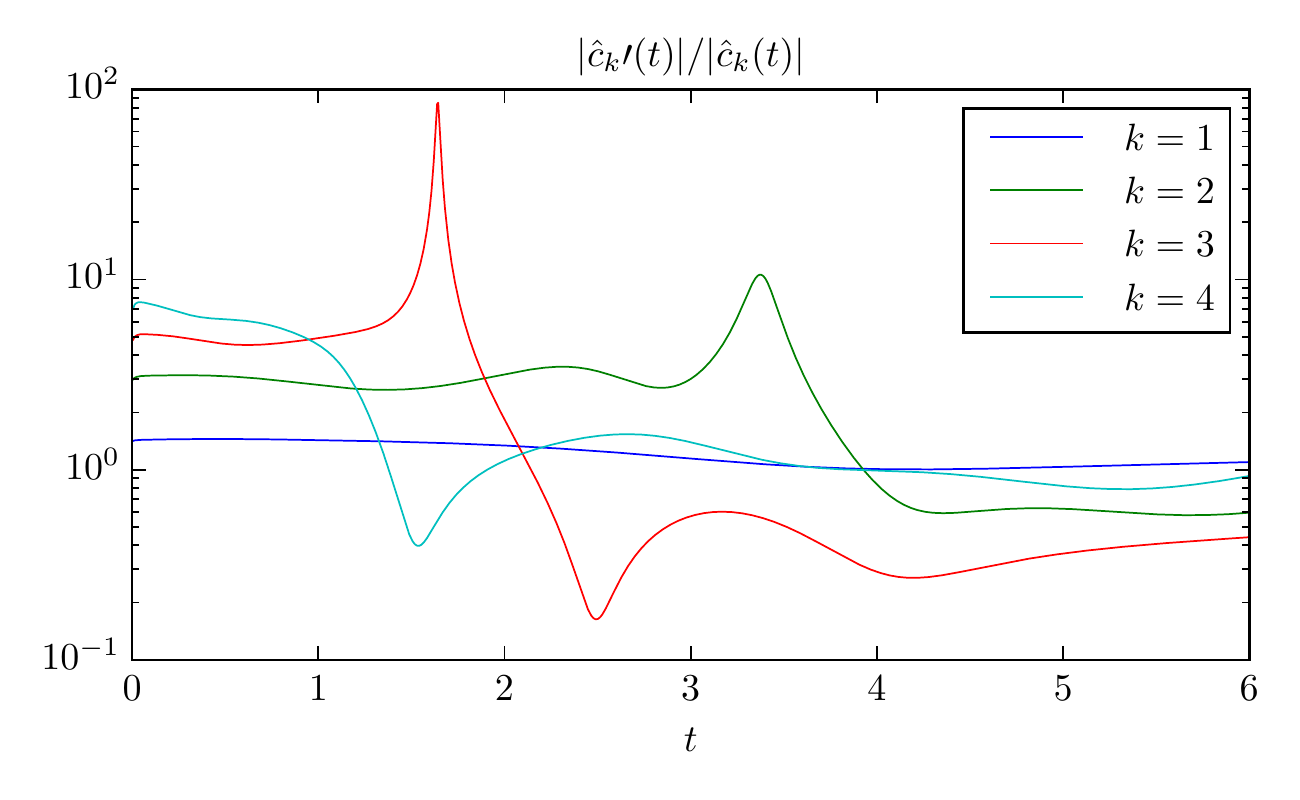}
    }\vspace{-.1cm}
\captionof{figure}{The log derivative varies significantly with time and over several orders of magnitude.}
		\label{fig:log_coeffs}
  \end{minipage}

\section{Conclusion}
In this paper a framework for defining an inverse problem to infer an infinite-dimensional linear operator was introduced by parametrizing the operator by its eigendecomposition.
Prior information about the behavior of the operator was included through deterministic structure imposed on its spectrum.
The operator in question for this research was assumed shift-invariant, which determined its eigenfunctions with certainty.
It was assumed a smoothing operator, which imposed bounds on the prior distributions of its eigenvalues.

Much more structure would need to be included in the operator parametrization to guarantee a physically relevant solution to the inverse problem.
The current formulation allows for operators that cause nonphysical solutions because it was not constrained to preserve positivity.
This can be an issue when there is not enough spatial data to constrain the solution to remain positive through the entire domain.
The challenge encountered in this study suggests that any and all prior knowledge is crucial to ensuring a successful calibration when data is sparse.

The operator inference was successful when using a large amount of spatial data spanning the entire domain.
Luckily, availability of a high-fidelity model makes this possible, and future work in developing the model inadequacy representation will make use of synthetic data.

The method for interrogating high-fidelity models proposed here has the potential to be a powerful tool in inadequacy model development.
It provides a way to directly inspect the result of incorrect assumptions on model error, which can guide the deterministic structure encoded in the inadequacy model.

\section{Future work}
To summarize, the goal of this work is to develop an inadequacy representation that characterizes the effects of using a closure model which fails to reproduce a physical phenomenon that significantly depends on missing fine-scale information.
To date, work has focused on building a framework for the Bayesian calibration of an operator and finding a means of understanding model error.

Future work will focus on interrogating the high-fidelity model by computing  ensemble averages over permeability fields for the evolution of the Fourier modes. 
This will help to discern how much the time-dependence of the log derivatives of the coefficients was caused by violating the shift-invariance assumption and how much is a result of losing depthwise information.

The evolution of model error can be computed by a direct comparison between the ensemble-averaged Fourier mode evolutions and the evolution predicted by the low-fidelity model (the 1D ADE in this case).
Information gleaned by inspecting this evolution can  be encoded in the structure of the inadequacy operator's eigenvalues by modeling their time evolution with a stochastic ODE.
This method of interrogation will prove equally useful in calibrating the stochastic inadequacy operator, because the inference can be split into independent inference problems for each eigenvalue using the ensemble average of each Fourier mode as calibration data.

The alternative approach of modeling the inadequacy as the solution to a stochastic PDE will be pursued simultaneously to compare the effectiveness of the two approaches. 
The approach that has been taken with some success in PECOS proceeds as follows:
an exact equation is derived for the model error, but it will have further unclosed terms.
These unclosed terms are replaced with a stochastic process to represent uncertainty in their values.

Finally, the methods developed for 1D low-fidelity model will be extended to develop an inadequacy representation for a 2D model of mean concentration resulting from an upscaling and averaging of a detailed 3D model.

\pagebreak
\bibliographystyle{plain}
\bibliography{MyBib}

\end{document}